# Feasible Proofs of Matrix Properties with Csanky's Algorithm

Michael Soltys

McMaster University
Computing and Software
1280 Main Street West
Hamilton, ON., Canada L8S 4K1
`soltys@mcmaster.ca`

**Abstract.** We show that Csanky's fast parallel algorithm for computing the characteristic polynomial of a matrix can be formalized in the logical theory **LAP**, and can be proved correct in **LAP** from the principle of linear independence. **LAP** is a natural theory for reasoning about linear algebra introduced in [8]. Further, we show that several principles of matrix algebra, such as linear independence or the Cayley-Hamilton Theorem, can be shown equivalent in the logical theory **QLA**. Applying the separation between complexity classes $\mathbf{AC}^0[2] \subsetneq \mathbf{DET}(\mathrm{GF}(2))$, we show that these principles are in fact not provable in **QLA**. In a nutshell, we show that linear independence is "all there is" to elementary linear algebra (from a proof complexity point of view), and furthermore, linear independence cannot be proved trivially (again, from a proof complexity point of view).

## 1 Introduction

This paper makes the following claim: our intuition that the principle of linear independence is all that there is to elementary linear algebra is justified from a proof complexity point of view. This means that from the principle of linear independence we can prove other strong principles of linear algebra (for example, the Cayley-Hamilton Theorem) using concepts of very low computational complexity. Furthermore, we claim that linear independence itself cannot be proved using concepts of low computational complexity.

To argue this claim, we present a new feasible proof of the Cayley-Hamilton Theorem (CHT) from the principle of linear independence in a weak theory of linear algebra (**QLAP**). The proof is based on Csanky's algorithm for computing the characteristic polynomial of a matrix. Csanky's algorithm is a fast parallel algorithm that computes the characteristic polynomial of a matrix over fields of characteristic zero.

**QLAP** is a first order theory for reasoning about matrices. Our new proof of the CHT with Csanky's algorithm leads to **QLAP** proofs of equivalence of important principles of linear algebra (for example, linear independence and the CHT). We also show that these principles are independent of **QLAP**. To show this independence we use the previously known result that $\mathbf{AC}^0[2]$ is properly contained in $\mathbf{DET}(\mathrm{GF}(2))$.

The class $\mathbf{AC}^0[2]$ consists of problems solvable with polynomial size circuits (in the size of the input), bounded depth, where besides the usual gates $\{\wedge, \vee, \neg\}$ we are also allowed to use the parity gate $\oplus$. The class $\mathbf{DET}(\mathrm{GF}(2))$ consists of problems $\mathbf{AC}^0$ reducible to computing the determinant over the field of two elements. Another class which will make a frequent appearance in this paper is $\mathbf{NC}^2$, which consists of those problems which are solvable with polynomial size circuits of depth $O(\log^2)$ (in the size of the input).

It is known that $\mathbf{AC}^0[2] \subsetneq \mathbf{DET}(\mathrm{GF}(2)) \subseteq \mathbf{NC}^2 \subseteq \mathbf{PolyTime}$, and the separation between the first two complexity classes is the famous result of Razborov and Smolensky ([5,6])). This separation will be instrumental in showing our independence result in the last section.

In this line of research we are motivated by a dual purpose: we want to understand the proof complexity of linear algebra, and we are also searching for good candidates for separating the Frege and extended Frege propositional proof systems. This separation is a central problem in theoretical computer science, and the theorems of universal linear algebra are considered to be good candidates to show such a separation—see [1] for more background on this quest.

In [8] we introduced the logical theory **LA** $\subset$ **LAP** $\subset$ ∃**LA** we gave the first feasible (i.e., using polynomial time concepts) proof of the CHT, a central theorem of matrix algebra from which many other universal theorems follow (in **LAP**). Our proof was based on Berkowitz's algorithm, which is an efficient parallel algorithm for computing the characteristic polynomial of a matrix (and hence the inverse, adjoint, and determinant of a matrix). Berkowitz's algorithm is field independent (that is, it works over any field), and it can be formalized with $\mathbf{NC}^2$ circuits. Both Berkowitz's algorithm and Csanky's algorithm are $\mathbf{NC}^2$ algorithms, and have the following interesting relationship: if they could be shown to compute the same thing in **LAP**, they could both be shown correct in **LAP**. As things stand now, are best proofs of correctness for both are polytime.

In section 2 we describe the relevant theories, **LA**, **LAP**, **QLA**, and ∃**LA**. In section 3 we describe Csanky's and Berkowitz's algorithms, and show that they can be formalized in **LAP**. In section 4 we show that the CHT follows in **LAP** from the principle of linear independence. This result is obtained using Csanky's algorithm, and so it requires fields of characteristic zero. In section 5 we show that five main principles of linear algebra can all be shown equivalent in **QLA**, and furthermore, **QLA** does not prove any of them.

## 2 The theories LA, LAP, ∃LA, and QLA

We define a quantifier-free theory of Linear Algebra (matrix algebra), and call it **LA**. Our theory is strong enough to prove the ring properties of matrices such as $A(BC) = (AB)C$ and $A + B = B + A$ but weak enough so that all the theorems of **LA** (over finite fields or the field of rationals) translate into propositional tautologies with short Frege proofs.

Our theory has three sorts of object: *indices* (i.e., natural numbers), *field elements*, and *matrices*, where the corresponding variables are denoted $i, j, k, \ldots$; $a, b, c, \ldots$; and $A, B, C, \ldots$, respectively. The semantics assumes that objects of type field are from a fixed but arbitrary field, and objects of type matrix have entries from that field.

Terms and formulas are built from the function and predicate symbols:

$$0_{\text{index}}, 1_{\text{index}}, +_{\text{index}}, *_{\text{index}}, -_{\text{index}}, \texttt{div}, \texttt{rem}, 0_{\text{field}}, 1_{\text{field}}, \\ +_{\text{field}}, *_{\text{field}}, -_{\text{field}}, {}^{-1}\texttt{r}, \texttt{c}, \texttt{e}, \Sigma, \leq_{\text{index}}, =_{\text{index}}, =_{\text{field}}, \\ =_{\text{matrix}}, \text{cond}_{\text{index}}, \text{cond}_{\text{field}} \quad (1)$$

The intended meanings should be clear, except for the following operations on a matrix $A$: $\texttt{r}(A), \texttt{c}(A)$ are the numbers of rows and columns in $A$, $\texttt{e}(A, i, j)$ is the field element $A_{ij}$, $\Sigma(A)$ is the sum of the elements in $A$. Also $\text{cond}(\alpha, t_1, t_2)$ is interpreted **if** $\alpha$ **then** $t_1$ **else** $t_2$, where $\alpha$ is a formula all of whose atomic subformulas have the form $m \leq n$ or $m = n$, where $m, n$ are terms of type index, and $t_1, t_2$ are terms either both of type index or both of type field. The subscripts $_{\text{index}}$ and $_{\text{field}}$ are usually omitted, since they are clear from the context.

In addition to the usual rules for constructing terms we also allow the terms $\lambda ij \langle m, n, t \rangle$ of type matrix. Here $i$ and $j$ are variables of type index bound by the $\lambda$ operator, intended to range over the rows and columns of the matrix. Here also $m, n$ are terms of type index *not* containing $i, j$ (representing the numbers of rows and columns of the matrix) and $t$ is a term of type field (representing the matrix element in position $(i, j)$).

The $\lambda$ terms allow us to construct the sum, product, transpose, etc., of matrices. For example, suppose first that $A$ and $B$ are $m \times n$ matrices. Then, $A + B$ can be defined as $\lambda ij \langle m, n, \texttt{e}(A, i, j) + \texttt{e}(B, i, j) \rangle$. Now suppose that $A$ and $B$ are $m \times p$ and $p \times n$ matrices, respectively. Then:

$$A * B := \lambda ij \langle m, n, \Sigma \lambda kl \langle p, 1, \texttt{e}(A, i, k) * \texttt{e}(B, k, j) \rangle \rangle$$

However, even if matrices are of incompatible size, their addition and product is well defined, since the "smaller" matrix is implicitly padded with zeros (as $\texttt{e}(A, i, j) = 0$ for $i$ or $j$ outside the range). Thus, all terms are well defined.

Atomic formulas and formulas are built in the usual manner, but in **LA** and **LAP** we only allow bounded index quantifiers (note that **LA**, respectively **LAP**, with bounded index quantifiers is conservative over **LA**, respectively **LAP**, without them).

We use Gentzen's sequent calculus LK (with quantifier rules omitted) for the underlying logic. We include 34 non-logical axioms in four groups: Axioms for equality, indices, field elements, and matrices (all quantifier-free). These specify the basic properties of the function and predicate symbols (1). By convention each instance of an axiom resulting from substituting terms for variables is also an axiom, so the axioms are really axiom schemes. All the axioms are given in [8].

We need an extra axiom to ensure that the underlying field is of characteristic zero. This can be stated with $\Sigma I_n \neq 0$, where $I_n$ is the $n \times n$ identity matrix, which is given with a constructed term $\lambda ij \langle n, n, \text{cond}(i = j, 1, 0) \rangle$. This requirement is necessary for Csanky's algorithm which works only over fields of characteristic zero, as it performs divisions by integers.

We need just two non-logical rules: an equality rule for terms of type matrix, and the induction rule:

$$\frac{\Gamma, \alpha(i) \to \alpha(i+1), \Delta}{\Gamma, \alpha(0) \to \alpha(n), \Delta} \quad (2)$$

To formalize Newton's and Berkowitz's algorithms we extend the theory **LA** to the theory **LAP** by adding a new function symbol $P$, where $P(n, A)$ means $A^n$. We also add two new axioms, which give a recursive definition of $P$; namely, $P(0, A) = I$ and $P(n + 1, A) = P(n, A) * A$. This is enough to formalize the coefficients of the characteristic polynomial of a matrix, as computed by either algorithm, as terms in the language of **LAP**. However, it seems that **LAP** is too weak to prove strong properties of the characteristic polynomial (such as the CHT or the multiplicativity of the determinant).

The theory $\exists$**LA** is an extension of **LA** where we allow induction over formulas of the form $(\exists X \leq t)\alpha$, where $\alpha$ has no quantifiers, and $\exists X \leq t$ is a bounded existential matrix quantifier ($X \leq t$ is just shorthand for $\mathtt{r}(X) \leq t \wedge \mathtt{c}(X) \leq t$). Note that the theory $\exists$**LAP**, defined analogously, is conservative over $\exists$**LA** because matrix powering ($P$) can be defined in $\exists$**LA**; so we don't really need to include $P$ (see [10]).

Finally, **QLA** is **LA** with quantification over matrices, but induction restricted to formulas of **LA**.

This concludes a brief tour through the theories **LA**, **LAP**, $\exists$**LA**, and **QLA**. They are natural theories, in that they include what one would expect to formalize matrix algebra. **LA** is the weakest, and it can be thought off as the theory that proves the ring properties of matrices. **LAP** is **LA** together with the matrix powering function (and defining axioms), and it can formalize Csanky's and Berkowitz's algorithm, but it seems too weak to prove strong properties about them. $\exists$**LA** is **LA** together with an induction over formulas with bounded matrix quantifiers (which also allows it to simulate **LAP**).

## 3 Csanky's and Berkowitz's algorithms

Both Csanky's and Berkowitz's algorithms compute the characteristic polynomial of a matrix, which is usually defined as $p_A(x) = \det(xI - A)$, for a given matrix $A$. Let $p_A^{\text{CSANKY}}$ and $p_A^{\text{BERK}}$ denote the coefficients of the characteristic polynomial of $A$ given as column vectors, respectively. Let $p_A^{\text{CSANKY}}(x)$ and $p_A^{\text{BERK}}(x)$ denote the actual characteristic polynomials, with coefficients computed by the respective algorithms.

Newton's symmetric polynomials are defined as follows: $s_0 = 1$, and for $1 \leq k \leq n$, by:

$$s_k = \frac{1}{k} \sum_{i=1}^{k} (-1)^{i-1} s_{k-i} \mathrm{tr}(A^i) \qquad (3)$$

Then, $p_A^{\mathrm{CSANKY}}(x) = s_0 x^n - s_1 x^{n-1} + s_2 x^{n-2} - \cdots \pm s_n x^0$. It is shown in the proof of lemma 1 how Csanky's algorithm computes the $s_i$'s more efficiently (in $\mathbf{NC}^2$) than in the straightforward way suggested by the recurrence (3).

**Lemma 1.** *$p_A^{\mathrm{CSANKY}}$ can be given as a term of $\boldsymbol{LAP}$.*

*Proof.* We follow the ideas in [11, Section 13.4]. We restate (3) in matrix form: $s = Ts - b$ where $s, T, b$ are given, respectively, as follows:

$$\begin{pmatrix} s_1 \\ s_2 \\ \vdots \\ s_n \end{pmatrix}, \quad \begin{pmatrix} 0 & 0 & 0 & \cdots \\ \frac{1}{2}\mathrm{tr}(A) & 0 & 0 & \cdots \\ \frac{1}{3}\mathrm{tr}(A^2) & \frac{1}{3}\mathrm{tr}(A) & 0 & \cdots \\ \frac{1}{4}\mathrm{tr}(A^3) & \frac{1}{4}\mathrm{tr}(A^2) & \frac{1}{4}\mathrm{tr}(A) & \cdots \\ \vdots & \vdots & \vdots & \ddots \end{pmatrix}, \quad \begin{pmatrix} \mathrm{tr}(A) \\ \frac{1}{2}\mathrm{tr}(A^2) \\ \vdots \\ \frac{1}{n}\mathrm{tr}(A^n) \end{pmatrix}$$

Then $s = -b(I - T)^{-1}$. Note that $(I - T)$ is an invertible matrix as it is lower triangular, with 1s on the main diagonal. The inverse of $(I - T)$ can be computed recursively using the following idea: if $C$ is lower-triangular, with no zeros on the main diagonal, then

$$C = \begin{pmatrix} C_1 & 0 \\ E & C_2 \end{pmatrix} \quad \Rightarrow \quad C^{-1} = \begin{pmatrix} C_1^{-1} & 0 \\ -C_2^{-1} E C_1^{-1} & C_2^{-1} \end{pmatrix}$$

There are $O(\log(n))$ many steps and the whole procedure can be simulated with circuits of depth $O(\log^2(n))$ and size polynomial in $n$.

This, however, does not give us an $\mathbf{LAP}$-term, and it would be difficult to formalize the proof of correctness of this recursive inversion procedure in $\mathbf{LAP}$. Thus, instead of this recursive computation, we use the fact that the CHT can be proved correct in $\mathbf{LAP}$ for *triangular* matrices (see [7, Section 5.2]). From the characteristic polynomial of $(I - T)$ we obtain its inverse, and the inverse can be proved correct (i.e., $(I - T)(I - T)^{-1} = (I - T)^{-1}(I - T) = I$) using the the CHT for triangular matrices, and this can be formalized in $\mathbf{LAP}$.

Berkowitz's algorithm, just as Csanky's algorithm, allows us to reduce the computation of the characteristic polynomial to matrix powering. Its advantage is that it works over any field; however, certain properties (such as the fact that similar matrices have the same characteristic polynomial) have easy proofs in weak theories ($\mathbf{LAP}$) for Csanky's algorithm, but (seem to) require polytime theories ($\exists \mathbf{LA}$) for Berkowitz's algorithm.

Berkowitz's algorithm computes the characteristic polynomial of a matrix in terms of the characteristic polynomial of its principal minor:

$$A = \begin{pmatrix} a_{11} & R \\ S & M \end{pmatrix} \tag{4}$$

where $R$ is an $1 \times (n-1)$ row matrix and $S$ is a $(n-1) \times 1$ column matrix and $M$ is $(n-1) \times (n-1)$. Let $p(x)$ and $q(x)$ be the characteristic polynomials of $A$ and $M$ respectively. Suppose that the coefficients of $p$ form the column vector

$$p = \begin{pmatrix} p_n & p_{n-1} & \ldots & p_0 \end{pmatrix}^t \tag{5}$$

where $p_i$ is the coefficient of $x^i$ in $\det(xI - A)$, and similarly for $q$. Then:

$$p = C_1 q \tag{6}$$

where $C_1$ is an $(n+1) \times n$ Toeplitz lower triangular matrix (Toeplitz means that the values on each diagonal are constant) and where the entries in the first column are defined as follows: $c_{i1} = 1$ if $i = 1$, $c_{i1} = -a_{11}$ if $i = 2$, and $c_{i1} = -(RM^{i-3}S)$ if $i \geq 3$. Berkowitz's algorithm consists in repeating this for $q$, and continuing so that $p$ is expressed as a product of matrices. Thus:

$$p_A^{\text{BERK}} = C_1 C_2 \cdots C_n \tag{7}$$

where $C_i$ is an $(n+2-i) \times (n+1-i)$ Toeplitz matrix defined as above except $A$ is replaced by its $i$-th principal sub-matrix. Note that $C_n = \begin{pmatrix} 1 & -a_{nn} \end{pmatrix}^t$.

Since each element of $C_i$ can be explicitly defined in terms of $A$ using matrix powering, and since the iterated matrix product can be reduced to matrix powering by a standard method, the entire product (7) can be expressed in terms of $A$ using matrix powering. Thus the right-hand side of (7) can be expressed as a term in **LAP**.

Since we can define the characteristic polynomial in **LAP** (as $p^{\text{CSANKY}}$ or $p^{\text{BERK}}$), it follows immediately that we can also define the determinant and the adjoint as terms of **LAP**.

## 4 Correctness of Csanky's Algorithm

The main result of this section, given as theorem 1, is the following:

$$\mathbf{QLAP} \vdash \text{Linear Independence} \supset \text{CHT} \tag{8}$$

where CHT (the Cayley-Hamilton Theorem) stands for $p_A(A) = p_A^{\text{CSANKY}}(A) = 0$. Since $\exists \mathbf{LA}$ proves the principle of linear independence (see [10]), we have a new proof that $\exists \mathbf{LA}$ can prove the CHT. We assume that the characteristic polynomial of $A$, $p_A$, is computed with Csanky's algorithm, i.e., in this section $p_A = p_A^{\text{CSANKY}}$.

**Lemma 2.** *LAP* proves that similar matrices have the same characteristic polynomial; that is, if $P$ is any invertible matrix, then $p_A = p_{PAP^{-1}}$.

*Proof.* Observe that $\text{tr}(AB) = \sum_i \sum_j a_{ij} b_{ji} = \sum_j \sum_i b_{ji} a_{ij} = \text{tr}(BA)$, so using the associativity of matrix multiplication, $\text{tr}(PA^i P^{-1}) = \text{tr}(A^i P P^{-1}) = \text{tr}(A^i)$. Inspecting (3), we see that a proof by induction on the $s_i$ proves this lemma.

**Lemma 3.** *LAP* proves that if $A$ is a matrix of the form:

$$\begin{pmatrix} B & 0 \\ C & D \end{pmatrix} \qquad (9)$$

where $B$ and $D$ are square matrices (not necessarily of the same size), and the upper-right corner is zero, then $p_A(x) = p_B(x) \cdot p_D(x)$.

*Proof.* Let $s_i^A, s_i^B, s_i^D$ be the coefficients of the characteristic polynomials (as given by (3)) of $A, B, D$, respectively. We want to show by induction on $i$ that

$$s_i^A = \sum_{j+k=i} s_j^B s_k^D,$$

from which the claim of the lemma follows. The Basis Case: $s_0^A = s_0^B = s_0^D = 1$. For the Induction Step, by definition and by the induction hypothesis, we have that $s_{i+1}^A$ equals

$$= \sum_{j=0}^{i} (-1)^j s_{i-j}^A \text{tr}(A^{j+1}) = \sum_{j=0}^{i} (-1)^j \left[ \sum_{p+q=i-j} s_p^B s_q^D \right] \text{tr}(A^{j+1})$$

and by the form of $A$ (i.e., (9)):

$$= \sum_{j=0}^{i} (-1)^j \left[ \sum_{p+q=i-j} s_p^B s_q^D \right] (\text{tr}(B^{j+1}) + \text{tr}(D^{j+1}))$$

to see how this formula simplifies, we divide it into two parts:

$$= \sum_{j=0}^{i} (-1)^j \left[ \sum_{p+q=i-j} s_p^B s_q^D \right] \text{tr}(B^{j+1}) + \sum_{j=0}^{i} (-1)^j \left[ \sum_{p+q=i-j} s_p^B s_q^D \right] \text{tr}(D^{j+1}).$$

Consider first the left-hand side. When $q = 0$, $p$ ranges over $\{i, i-1, \ldots, 0\}$, and $j+1$ ranges over $\{1, 2, \ldots, i+1\}$, and therefore, by definition, we obtain $s_{i+1}^B$. Similarly, when $q = 1$, we obtain $s_i^B$, and so on, until we obtain $s_1^B$. Hence we have:

$$= \sum_{j=0}^{i+1} s_{i-j}^B s_j^D + \sum_{j=0}^{i} (-1)^j \left[ \sum_{p+q=i-j} s_p^B s_q^D \right] \text{tr}(D^{j+1}).$$

The same reasoning, but fixing $p$ instead of $q$ on the right-hand side, gives us:

$$= \sum_{j=0}^{i+1} s_{i-j}^B s_j^D + \sum_{j=0}^{i+1} s_j^B s_{i-j}^D = \sum_{j+k=i+1} s_j^B s_k^D$$

which gives us the induction step and the proof of the lemma.

To show that $p_A(A) = 0$ it is sufficient to show that $p_A(A)e_i = 0$ for all vectors $e_i$ in the standard basis $\{e_1, e_2, \ldots, e_n\}$. Let $k$ be the largest integer such that

$$\{e_i, Ae_i, \ldots, A^{k-1}e_i\} \qquad (10)$$

is linearly independent; we know that $k - 1 < n$, by the principle of linear independence (this is the first place where we use linear independence). Then, (10) is a basis for a subspace $W$ of $\mathbb{F}^n$, and $W$ is invariant under $A$, i.e., given any $w \in W$, $Aw \in W$.

Using Gaussian Elimination we write $A^k e_i$ as a linear combination of the vectors in (10). Using the coefficients of this linear combination we write a monic polynomial

$$g(x) = x^k + c_1 x^{k-1} + \cdots + c_k x^0 \qquad (11)$$

such that $g(A)e_i = 0$.

Let $A_W$ be $A$ restricted to the basis (10), that is, $A_W$ is a matrix representing the linear transformation $T_A : \mathbb{F}^n \longrightarrow \mathbb{F}^n$ induced by $A$, restricted to the subspace $W$. The matrix $A_W^t$ has the following simple form:

$$\begin{pmatrix} 0 & 0 & 0 & \ldots & 0 & -c_k \\ 1 & 0 & 0 & \ldots & 0 & -c_{k-1} \\ 0 & 1 & 0 & \ldots & 0 & -c_{k-2} \\ \vdots & & \ddots & & & \vdots \\ 0 & 0 & 0 & \ldots & 1 & -c_1 \end{pmatrix} \qquad (12)$$

i.e., it is the *companion matrix* of the polynomial $g(x)$. Since $p_A = p_{A^t}$, we consider the transpose of $A_W$, since $A_W^t$ has the property that its principal submatrix is also a companion matrix, and that will be used in a proof by induction in the next lemma.

The proof of the next lemma is the crucial technical result of this section. The proof is given in the appendix.

**Lemma 4.** ***LAP*** *proves that the polynomial* $g(x)$ *is the characteristic polynomial of* $A_W$, *in other words,* $g(x) = p_{A_W}(x)$.

It is interesting to note that lemma 4 can also be proved (feasibly) for Berkowitz's algorithm instead, and the proof is in fact much simpler: consider again the matrix given by (12). We assume inductively that $p_M^{\text{BERK}}$ (the characteristic polynomial of the principal submatrix of (12)) is given by $(1\ c_1\ c_2\ \ldots\ c_{k-1})^t$. Since $R = (0 \ldots 0\ -c_k)$ and $S = e_1$, $p_A^{\text{BERK}} = B \cdot p_M^{\text{BERK}}$, where $B$ (the matrix given by Berkowitz's algorithm) is an $(n+1) \times n$ matrix with 1s on the main diagonal, 0s everywhere else, except for $+c_k$ in position $(n+1,1)$. From this, it is easy to see that $p_A^{\text{BERK}}$ is given by $(\, 1 \; c_1 \; c_2 \; \ldots \; c_k \,)^t$.

As was pointed out in the introduction, if we managed to prove in **LAP** that Csanky's and Berkowitz's algorithms compute the same thing (i.e., $p^{\text{CSANKY}} = p^{\text{BERK}}$) we would have an **LAP** proof of the CHT for both. The reason is that the CHT follows for Berkowitz's algorithm from $\det(A) = \det(PAP^{-1})$, which is trivial to prove for Csanky's algorithm (see proof of Lemma 2).

**Lemma 5.** $\exists\mathbf{LA}$ *proves that the polynomial $g(x)$ divides $p_A(x)$.*

*Proof.* Extend (10) to a full basis of $\mathbb{F}^n$:

$$B = \{e_i, Ae_i, \ldots, A^{k-1}e_i, e_{j_1}, e_{j_2}, \ldots, e_{j_{n-k}}\}.$$

This extension can be carried out easily with Gaussian Elimination, by checking which vectors from the standard basis ($\{e_1, e_2, \ldots, e_n\}$) are in the span consisting of (10) and those vectors that have already been added, and adding only those that are not. This is the only other place (besides the paragraph following the proof of lemma 3) where we need to use the principle of linear independence.

Let $P$ be the change of basis for $A$ from the standard basis to $B$. Then,

$$PAP^{-1} = \begin{pmatrix} A_W & 0 \\ * & E \end{pmatrix}$$

where $A_W$ is a $k \times k$ block, and $E$ is a $(n-k) \times (k-n)$ block (corresponding to the extension), and we have a block of zeros above $E$ since $W$ is invariant under $A$. By lemma 3 it follows that $p_A(x) = p_{PAP^{-1}}(x) = p_{A_W}(x) \cdot p_E(x)$. By lemma 4, $p_{A_W} = g(x)$, and so $g(x)$ divides $p_A(x)$.

**Theorem 1.** *$\mathbf{QLAP}$ proves the Cayley-Hamilton Theorem (CHT) from the principle of linear independence, when the characteristic polynomial is computed by Csanky's algorithm.*

*Proof.* By lemma 5,

$$p_A(A)e_i = (p_{A_W}(A) \cdot p_E(A))e_i = (g(A) \cdot p_E(A))e_i = p_E(A) \cdot (g(A)e_i) = 0.$$

Since this is true for any $e_i$ in the standard basis, it follows that $p_A(A) = 0$.

The proof of the multiplicativity of the determinant is a $\exists\mathbf{LA}$ corollary of this theorem, as can be seen in [8]. Together, the CHT and the multiplicativity of the determinant, are two powerful universal principles of linear algebra from which many others follow directly. An important open question remains: are they provable in **LAP**?

## 5 Equivalence of Matrix Principles

Consider the following five central principles of linear algebra:

1. The Cayley-Hamilton Theorem
2. $(\exists B \neq 0)[AB = I \vee AB = 0]$
3. Linear Independence ($n+1$ vectors in $\mathbb{F}^n$ must be linearly dependent)
4. Weak Linear Independence ($n^k$ vectors ($n, k > 1$) in $\mathbb{F}^n$ must be linearly dependent)
5. Every matrix has an annihilating polynomial

In this section we are going to show that **QLA** proves their equivalence. Furthermore, we show that these principles are *independent* of **QLA**. Thus, even though **QLA** is strong enough to show them equivalent, it is too weak to prove any of them.

Notice however that **QLA** does not have the matrix powering function, yet two of these principles, namely 1 and 5, require matrix powering to be stated. Let $\text{POW}(A, n)$ be the formula:

$$\exists \langle X_0 X_1 \ldots X_n \rangle (\forall i \leq n)[X_0 = I \wedge (i < n \supset X_{i+1} = X_i * A)] \quad (13)$$

The size of $\langle X_0 X_1 \ldots X_n \rangle$ can be bounded as it is a $\mathtt{r}(A) \times (\mathtt{r}(A) \cdot (n+1))$ matrix. (The abuse of notation in (13) is for better readability, but this formula can be stated formally as a bounded $\Sigma_1$ formula of **QLA**.)

**Theorem 2.** *The five principles of linear algebra can be proved equivalent in **QLAP** with $\text{POW}(A, n)$.*

*Proof.* 3 implies 1 because of the results of the previous section. Note that here we need fields of characteristic zero (because of Csanky's algorithm). It is an open question whether we can prove this over arbitrary fields—for example in the context of Berkowitz's algorithm.

1 implies 2 because $B$ is just the adjoint, for which we have the desired properties from the Cayley-Hamilton Theorem.

2 implies 3, because suppose that we have $(n+1)$ vectors in $\mathbb{F}^n$, and that they are linearly independent. Let $A$ be the $n \times (n+1)$ matrix whose columns are these $n+1$ vectors. Let $A'$ be the matrix resulting by appending a row of zeros to $A$. Since the vectors are linearly independent, there is no $B$ such that $A'B = 0$, so by 2 there must be a $B$ such that $A'B = I$; but that is not possible, given that the last row of $A'$ is zero.

3 obviously implies 4.

4 implies 5 because we can look at $\{I, A, A^2, \ldots, A^{n^k}\}$, where $A$ is $n \times n$, and $k$ as large as we want, and as vectors these matrices are linearly dependent by 4.

5 implies 2, because if $p(A) = 0$, we can choose the largest $s$ such that $p(A) = q(A)A^s$. If $q(A) \neq 0$, we choose the largest $k \leq s$ so that $q(A)A^k \neq 0$, and this is our zero divisor for $A$. If $q(A) = 0$, then it has a non-zero constant coefficient, and hence we can obtain from $q(A)$ the inverse for $A$.

Recall that the **Steinitz Exchange Theorem (SET)** says the following: if $T$ is a (finite) total set for a vector space $V$, i.e., $\text{span}(T) = V$, and $E$ is a *linearly independent* set, then there exists an $F \subseteq T$, such that $|F| = |E|$, and $(T - F) \cup E$ is total. (Note that in general, SET is stated for any $T$, not necessarily finite, but here we assume that $T$ is finite.)

We can state SET in the language of **QLA** as follows: associate the finite set $T$ of $m$ vectors in $\mathbb{F}^n$ with a $n \times m$ matrix $T$, and we can state that $T$ is total with $(\exists A)[TA = I]$. Let $E$ be a $n \times k$ matrix representing the $k$ vectors in $E$. We want to find $k$ column in $T$, and replace them by $E$. We can state that there exists a permutation matrix so that $TP$ has those $k$ columns as the last $k$ columns. Using the $\lambda$-constructor, we can "chop of" those last $k$ columns, and replace them by $E$, and then state that the result is also total. Thus, SET can be stated in **QLA**.

**Lemma 6.** *QLA proves that the Steinitz Exchange Theorem implies the five principles listed at the beginning of this section.*

*Proof.* We show that SET implies (in **QLA**) the existence of an annihilating polynomial. Consider the set $E = \{I, A, A^2, A^3, \ldots, A^{n^2-1}\}$, where $A$ is an $n \times n$ matrix. If $E$ is linearly dependent, we are done: we have an annihilating polynomial. Otherwise, suppose that $E$ is linearly independent.

Let $V = M_{n \times n}(\mathbb{F})$, that is $V$ is the vector space of $n \times n$ matrices, over some field $\mathbb{F}$ (note that our argument is field independent). Let $T = \{e_{ij}\}_{1 \leq i,j \leq n}$, that is, $T$ is the set of all elementary matrices $e_{ij}$, which are matrices with 1 in position $(i,j)$, but zeros everywhere else. Note that $|T| = |E| = n^2$, and $T$ is clearly total.

Therefore, by the Steinitz Exchange Theorem, $(T - F) \cup E$ is total for some $|F| = |E|$, and so $E$ is total since $T = F$ if $|T| = |E| = n^2$. If $E$ is total, then $A^{n^2} \in \text{span}(E)$, and hence $E \cup \{A^{n^2}\}$ is linearly dependent, and so we have an annihilating polynomial once again.

Can we show that the five principles, listed at the beginning of this section, prove (in **QLA**) the SET? Here is an obvious proof of SET: pick $E_1$ in $E$, and since $T$ is total, we can write it as a linear combination of elements in $T$, say $E_1 = a_1 T_1 + a_2 T_2 + \cdots a_n T_n$, all $a_i \neq 0$. So, $T_1$ can be written as a sum of elements in $T - \{T_1\} \cup \{E_1\}$. So, put $T_1$ in $F$. Note that $T - \{T_1\} \cup \{E_1\}$ remains total. Now pick $E_2$, and write it as a linear combination of a finite subset of elements in $T - \{T_1\} \cup \{E_1\}$. By the assumed linear independence of $E$, $E_2$ cannot be written in terms of $E_1$ alone, so like before, we can pick some $T_2$ and put it in $F$. We proceed inductively, at each step putting some $T_i$ in $F$.

The problem with the proof outlined above is that it requires induction over formulas with matrix quantifiers, which we do not have in **QLA** (on the other hand, this proof could be easily formalized in $\exists$**LA**). Thus we propose the following open problem: can SET be proved in **QLA** from the five principles? More generally: can Gaussian Elimination, properly stated, be shown correct in **QLA** from the five principles?

We conjecture that the answer is "yes" to those two questions, and that they are not too hard to prove.

**Lemma 7.** $\mathbf{QLA} \vdash (\exists B \neq 0)[AB = I \vee AB = 0] \supset POW(A, n)$.

*Proof.* We use reduction of matrix powering to matrix inverse described in [3]. Let $N$ be the $n^2 \times n^2$ matrix consisting of $n \times n$ blocks which are all zero except for $(n-1)$ copies of $A$ above the diagonal zero blocks. Then $N^n = 0$, and $(I - N)^{-1} = I + N + N^2 + \ldots + N^{n-1} =$

$$\begin{pmatrix} I & A & A^2 & \ldots & A^{n-1} \\ 0 & I & A & \ldots & A^{n-2} \\ \vdots & & & \ddots & \vdots \\ 0 & 0 & 0 & \ldots & I \end{pmatrix}.$$

Set $C = I - N$. Show that if $CB = 0$, then $B = 0$, using induction on the rows of $B$, starting with the bottom row. Using $(\exists B \neq 0)[CB = I \vee CB = 0]$, conclude that there is a $B$ such that $CB = I$. Next, show that $B = I + N + N^2 + \cdots + N^{n-1}$, again, by induction on the rows of $B$, starting with the bottom row. Thus, $B$ contains $I, A, A^2, \ldots, A^{n-1}$ in its top rows, and $POW(A, n)$ follows.

Thus, not every implication in theorem 2 requires $POW(A, n)$. In particular, $2 \Leftrightarrow 3$ and $3 \Rightarrow 4$ can be shown in **QLA** (for $2 \Leftrightarrow 3$ see proof of corollary below). It is an open question whether 4 implies 3 in **QLA**.

**Lemma 8.** $\mathbf{QLA} \nvdash POW(A, n)$.

*Proof.* We can turn **QLA** into a three-sorted universal theory in the style of **QPV** ([2]), by introducing function symbols for all the $\lambda$-terms, so we have number-valued functions, field-valued functions, and matrix valued-functions. Further, if the underlying field is $GF(2)$, then all these functions are in the complexity class $\mathbf{AC}^0[2]$ (by translations given in [8]). Hence, by the Herbrand Theorem, every existential theorem of **QLA** can be witnessed by an $\mathbf{AC}^0[2]$ function.

Let $\mathbf{DET}(GF(2))$ be the complexity class of functions $\mathbf{NC}^1$ reducible to the determinant over $GF(2)$. This class is equal to the class $\mathbf{POW}(GF(2))$, by results in [3]. On the other hand, $\mathbf{AC}^0[2]$ is properly contained in $\mathbf{DET}(GF(2))$, since $\mathbf{L} \subseteq \mathbf{DET}(GF(2))$ (see [4]), while $MAJORITY \in \mathbf{L}$ but it is *not* in $\mathbf{AC}^0[2]$ (see [5,6]).

**Corollary 1.** *$\mathbf{QLA}$ does not prove the principles 2 and 3 (while it can show them equivalent without $POW(A, n)$).*

*Proof.* By lemmas 7 and 8 we see that **QLA** does not prove 2. Now, 3 implies 2 by the following argument: take $A$ and add $e_i$ (the elementary column vector with 1 in the $i$-th entry, and zeros everywhere else) as the last column. By linear independence, we know that there exist $b_{1i}, b_{2i}, \ldots, b_{(n+1)i}$, not all zero, such that $b_{1i}A_1 + b_{2i}A_2 + \cdots b_{ni}A_n + b_{(n+1)i}e_i = 0$, where $A_i$ is the $i$-th column of $A$. If for all $i$, $b_{(n+1)i}$ is not zero, we found $B$ such that $AB = I$. If, on the other hand, some $b_{(n+1)i} = 0$, then $B$ consisting of columns given by $[b_{1i}b_{2i}\ldots b_{ni}]^t$ is a zero divisor of $A$, i.e., $AB = 0$.

## 6   Conclusions and Open Problems

We gave a new feasible proof of the Cayley-Hamilton Theorem via Csanky's algorithm. The new proof requires fields of characteristic zero, but it shows that the CHT follows in **LAP** from the principle of linear independence. It is an open question whether the CHT follows in **LAP** from the principle of linear independence over general fields.

We showed that five important principles of linear algebra can be shown equivalent in **QLA**, and using a previously known separation of complexity classes (namely $\mathbf{AC}^0[2] \subsetneq \mathbf{DET}(\mathrm{GF}(2))$) we showed that none of these principles is provable in **QLA**.

It is an interesting open problem whether the principles listed in theorem 2 can be proved in $\mathbf{QLA} + \mathrm{POW}(A, n)$. Likewise, it is an open problem whether Berkowitz's and Csanky's algorithm are provable correct in **LAP** (they can be stated in **LAP**, and weak properties of correctness are provable in **LAP**).

**Acknowledgments:** The author would like to thank Stephen Cook for pointing out the proof of the Cayley-Hamilton Theorem in [9], which is the basis for the proof in section 4. The material in section 5 came from discussions with Mark Braverman and Stephen Cook. Finally, the author is grateful to the anonymous referees, especially to the referee who succinctly and elegantly expressed the contribution of this paper (see the first sentence of the introduction).

## 7 Appendix

*Proof (lemma 4).* We will drop the $W$ from $A_W$ as there is no danger of confusion (the original matrix $A$ does not appear in the proof); thus, $A$ is a $k \times k$ matrix, with 1s below the main diagonal, and zeros everywhere else except (possibly) in the last column where it has the negations of the coefficients of $g(x)$.

As was noted above, $A$ is divided into four quadrants, with the upper-left containing just 0. Let $R = (0 \ldots 0 \; -c_k)$ be the row vector in the upper-right quadrant. Let $S = e_1$ be the column vector in the lower-left quadrant, i.e., the first column of $A$ without the top entry. Finally, let $M$ be the principal submatrix of $A$, $M = A[1|1]$; the lower-right quadrant.

Let $s_0, s_1, \ldots, s_k$ be the Newton's symmetric polynomials of $A$.

To prove that $g(x) = p_{A_{T_W}}(x)$ we prove something stronger: we show that (i) for all $0 \leq i \leq k$ $(-1)^i s_i = c_i$, and (ii) $p_A(A) = 0$.

We show this by induction on the size of the matrix $A$. Since the principal submatrix of $A$ (i.e., $M$) is *also* a companion matrix, we assume that for $i < k$, the coefficients of the symmetric polynomial of $M$ are equal to the $c_i$'s, and that $p_M(M) = 0$. (Note that the Basis Case of the induction is a $1 \times 1$ matrix, and it is trivial to prove.)

Since for $i < k$, $\mathrm{tr}(A^i) = \mathrm{tr}(M^i)$, it follows from (3) and the induction hypothesis that for $i < k$, $(-1)^i s_i = c_i$ (note that $s_0 = c_0 = 1$).

Next we show that $(-1)^k s_k = c_k$. By definition (i.e., by (3)) we have that $s_k$ is equal to:

$$\frac{1}{k}(s_{k-1}\mathrm{tr}(A) - s_{k-2}\mathrm{tr}(A^2) + \cdots + (-1)^{k-2}s_1\mathrm{tr}(A^{k-1}) + (-1)^{k-1}s_0\mathrm{tr}(A^k))$$

and by the induction hypothesis and the fact that for $i < k$ $\mathrm{tr}(A^i) = \mathrm{tr}(M^i)$ we have:

$$= \frac{1}{k}(-1)^{k-1}(c_{k-1}\mathrm{tr}(M) + c_{k-2}\mathrm{tr}(M^2) + \cdots + c_1\mathrm{tr}(M^{k-1}) + c_0\mathrm{tr}(A^k)).$$

Note that $\mathrm{tr}(A^k) = -kc_k + \mathrm{tr}(M^k)$, so:

$$= \frac{1}{k}(-1)^{k-1}\left[c_{k-1}\mathrm{tr}(M) + c_{k-2}\mathrm{tr}(M^2) + \cdots + c_1\mathrm{tr}(M^{k-1}) + c_0\mathrm{tr}(M^k)\right]$$
$$+ (-1)^k c_k$$

Observe that

$$\mathrm{tr}(c_{k-1}M + c_{k-2}M^2 + \cdots + c_1 M^{k-1} + c_0 M^k) = \mathrm{tr}(p_M(M)M) = \mathrm{tr}(0) = 0$$

since $p_M(M) = 0$ by the induction hypothesis. Therefore, $s_k = (-1)^k c_k$.

It remains to prove that $p_A(A) = \sum_{i=0}^{k} c_i A^{k-i} = 0$. First, show that for $1 \leq i \leq (k-1)$:

$$A^{i+1} = \left( \begin{array}{c|c} 0 & RM^i \\ \hline M^i S & \sum_{j=0}^{i-1} M^j SRM^{(i-1)-j} + M^{i+1} \end{array} \right) \tag{14}$$

(For $A$ of the form given by (12), and $R, S, M$ defined as in the first paragraph of the proof.) Define $w_i, X_i, Y_i, Z_i$ as follows:

$$\begin{aligned} A^{i+1} &= \begin{pmatrix} w_{i+1} & X_{i+1} \\ Y_{i+1} & Z_{i+1} \end{pmatrix} = \begin{pmatrix} w_i & X_i \\ Y_i & Z_i \end{pmatrix} \begin{pmatrix} 0 & R \\ S & M \end{pmatrix} \\ &= \begin{pmatrix} X_i S & w_i R + X_i M \\ Z_i S & Y_i R + Z_i M \end{pmatrix} \end{aligned} \tag{15}$$

We want to show that the right-most matrix of (15) is equal to the right-hand side of (14). First note that:

$$X_{i+1} = \sum_{j=0}^{i} w_{i-j} RM^j \qquad w_{i+1} = \sum_{j=0}^{i-1} (RM^j S) w_{i-1-j} \tag{16}$$

With the convention that $w_0 = 1$. See [8, lemma 5.1] for an **LAP**-proof of (16). Since $w_1 = 0$, a straight-forward induction shows that $w_{i+1} = 0$. Therefore, at this point the right-most matrix of (15) can be simplified to:

$$\begin{pmatrix} 0 & RM^i \\ Z_i S & Y_i R + Z_i M \end{pmatrix}$$

Again by [8, lemma 5.1] we have:

$$Y_{i+1} = M^i S + \sum_{j=0}^{i-2} (RM^j S) Y_{i-1-j} \qquad Z_{i+1} = M^{i+1} + \sum_{j=0}^{i-1} Y_{i-1-j} RM^j$$

By the same reasoning as above, $\sum_{j=0}^{i-2}(RM^j S)Y_{i-1-j} = 0$, so putting it all together we obtain the right-hand side of (14).

Using the induction hypothesis ($p_M(M) = 0$) it is easy to show that the first row and column of $p_A(A)$ are zero. Also, by the induction hypothesis, the term $M^{i+1}$ in the principal submatrix of $p_A(A)$ disappears but leaves $c_k I$. Therefore, it will follow that $p_A(A) = 0$ if we show that

$$\sum_{i=2}^{k} c_{k-i} \sum_{j=0}^{i-2} M^j SRM^{(i-2)-j} \tag{17}$$

is equal to $-c_k I$.

Some observations about (17): for $0 \leq j \leq i-2 \leq k-2$, the first column of $M^j$ is just $e_{j+1}$. And $SR$ is a matrix of zeros, with $-c_k$ in the upper-right corner. Thus $M^j SR$ is a matrix of zeros except for the last column which is $-c_k e_{j+1}$. Thus, $M^j SR M^{(i-2)-j}$ is a matrix with zeros everywhere, except in row $(j+1)$ where it has the bottom row of $M^{(i-2)-j}$ multiplied by $-c_k$. Let $\mathbf{m}^{(i-2)-j}$ denote the $1 \times (k-1)$ row vector consisting of the bottom row of $M^{(i-2)-j}$. Therefore, (17) is equal to:

$$-c_k \cdot \begin{pmatrix} \sum_{i=2}^{k} c_{k-i}\mathbf{m}^{(i-2)} \\ \hline \sum_{i=3}^{k} c_{k-i}\mathbf{m}^{(i-3)} \\ \hline \vdots \\ \hline \sum_{i=k}^{k} c_{k-i}\mathbf{m}^{(i-k)} \end{pmatrix} \quad (18)$$

We want to show that (18) is equal to $-c_k I$ to finish the proof of $p_A(A) = 0$. To accomplish this, let $l$ denote the $l$-th row of the matrix in (18) starting with the bottom row. We want to show, by induction on $l$, that the $l$-th row is equal to $e_{k-l}$.

The Basis Case is $l = 0$:

$$\sum_{i=k}^{k} c_{k-i}\mathbf{m}^{(i-k)} = c_0 \mathbf{m}^0 = e_k,$$

and we are done.

For the induction step, note that $\mathbf{m}^{l+1}$ is equal to $\mathbf{m}^l$ shifted to the left by one position, and with

$$\mathbf{m}^l \cdot (\,-c_{k-1}\ -c_{k-2}\ \ldots\ -c_1\,)^t \quad (19)$$

in the last position. We introduce some more notation: let $\mathbf{r}_l$ denote the $k-l$ row of (18). Thus $\mathbf{r}_l$ is $1 \times (k-1)$ row vector. Let $\overleftarrow{\mathbf{r}}_l$ denote $\mathbf{r}_l$ shifted by one position to the left, and with a zero in the last position. This can be stated succinctly in **LAP** as follows:

$$\overleftarrow{\mathbf{r}}_l \stackrel{\text{def}}{=} \lambda ij\langle 1, (k-1), e(\mathbf{r}_l, 1, i+1))\rangle.$$

Based on (18) and (19) we can see that:

$$\mathbf{r}_{l+1} = \overleftarrow{\mathbf{r}}_l + [\mathbf{r}_l \cdot (\,-c_{k-1}\ -c_{k-2}\ \ldots\ -c_1\,)^t] e_k + c_l \mathbf{m}^0.$$

(Here the "·" in the square brackets denotes the dot product of the two vectors.) Using the induction hypothesis: $\overleftarrow{\mathbf{r}}_l = e_{k-(l+1)}$, and

$$\mathbf{r}_l \cdot (\,-c_{k-1}\ -c_{k-2}\ \ldots\ -c_1\,)^t = e_{k-l} \cdot (\,-c_{k-1}\ -c_{k-2}\ \ldots\ -c_1\,)^t = -c_l$$

so $\mathbf{r}_{l+1} = e_{k-l} - c_l e_k + c_l e_k = e_{k-(l+1)}$ as desired. This finishes the proof of the fact that the matrix in (18) is the identity matrix, which in turn proves that (17) is equal to $-c_k I$, and this ends the proof of $p_A(A) = 0$, which finally finishes the main induction argument, and proves the lemma.